\documentclass[10pt]{nature1mio}

\usepackage{graphicx,latexsym,amssymb,multirow,bigstrut}
\newcommand {\hess} {H.E.S.S. }

\newcommand {\C}{Cherenkov }

\newcommand {\m} {$\rm \mu m$}
\newcommand {\rchisq} {$\chi_{\rm red}^{2}$}

\newcommand {\nw} {nW m$^{-2}$ sr$^{-1}$}

\newcommand {\fu}{cm$^{-2}$s$^{-1}$TeV$^{-1}$}

\title{A low level of extragalactic background light \\
as revealed by $\mathbf \gamma$-rays from blazars}

\author{F. Aharonian$^{1}$, 
 A.G.~Akhperjanian$^{2}$, 
 A.R.~Bazer-Bachi$^{3}$, 
 M.~Beilicke$^{4}$, 
 W.~Benbow$^{1}$, 
 D.~Berge$^{1}$, 
 K.~Bernl\"ohr$^{1,5}$, 
 C.~Boisson$^{6}$, 
 O.~Bolz$^{1}$, 
 V.~Borrel$^{3}$
% ...
}

\date{ }

\begin{document}
\maketitle


\begin{affiliations}
\small
\item Max-Planck-Institut f\"ur Kernphysik, P.O. Box 103980, D 69029
Heidelberg, Germany
\item Yerevan Physics Institute, 2 Alikhanian Brothers St., 375036 Yerevan,
Armenia
\item Centre d'Etude Spatiale des Rayonnements, CNRS/UPS, 9 av. du Colonel Roche, BP
4346, F-31029 Toulouse Cedex 4, France
\item Universit\"at Hamburg, Institut f\"ur Experimentalphysik, Luruper Chaussee
149, D 22761 Hamburg, Germany
\item Institut f\"ur Physik, Humboldt-Universit\"at zu Berlin, Newtonstr. 15,
D 12489 Berlin, Germany
\item LUTH, UMR 8102 du CNRS, Observatoire de Paris, Section de Meudon, F-92195 Meudon Cedex,
France
\item University of Durham, Department of Physics, South Road, Durham DH1 3LE,
U.K.
\item Laboratoire Leprince-Ringuet, IN2P3/CNRS,
Ecole Polytechnique, F-91128 Palaiseau, France
\item
APC, 11 Place Marcelin Berthelot, F-75231 Paris Cedex 05, France 
\item Dublin Institute for Advanced Studies, 5 Merrion Square, Dublin 2,
Ireland
\item
Landessternwarte, K\"onigstuhl, D 69117 Heidelberg, Germany
\item
Laboratoire de Physique Th\'eorique et Astroparticules, IN2P3/CNRS,
Universit\'e Montpellier II, CC 70, Place Eug\`ene Bataillon, F-34095
Montpellier Cedex 5, France
\item
Laboratoire d'Astrophysique de Grenoble, INSU/CNRS, Universit\'e Joseph Fourier, BP
53, F-38041 Grenoble Cedex 9, France 
\item
DAPNIA/DSM/CEA, CE Saclay, F-91191
Gif-sur-Yvette, Cedex, France
\item
Unit for Space Physics, North-West University, Potchefstroom 2520,
    South Africa
\item
Laboratoire de Physique Nucl\'eaire et de Hautes Energies, IN2P3/CNRS, Universit\'es
Paris VI \& VII, 4 Place Jussieu, F-75252 Paris Cedex 5, France
\item
Institut f\"ur Theoretische Physik, Lehrstuhl IV: Weltraum und
Astrophysik,
    Ruhr-Universit\"at Bochum, D 44780 Bochum, Germany
\item
Institute of Particle and Nuclear Physics, Charles University,
    V Holesovickach 2, 180 00 Prague 8, Czech Republic
\item
University of Namibia, Private Bag 13301, Windhoek, Namibia
\item
European Associated Laboratory for Gamma-Ray Astronomy, jointly
supported by CNRS and MPG
\end{affiliations}

\vskip 0.5cm

\begin{abstract}
The diffuse extragalactic background light consists of the sum of the 
starlight emitted by galaxies through the history of the Universe, 
and it could also have an important contribution from the  first stars, 
which may have formed before galaxy formation began. 
Direct measurements are difficult and not yet conclusive, 
owing to the large uncertainties caused by the bright foreground emission 
associated with zodiacal light\cite{hauser}. 
An alternative approach\cite{nik,gould,jelly,steck92} is to study 
the absorption features imprinted on the $\gamma$-ray spectra of distant 
extragalactic objects by interactions of those photons with the background 
light photons\cite{aha01}.
Here we report the discovery of $\gamma$-ray emission from 
the blazars\cite{costagg} H\,2356$-$309 and 1ES\,1101$-$232, 
at redshifts z=0.165 and z=0.186, respectively.
Their unexpectedly hard spectra provide an upper limit on the 
background light at optical/near-infrared wavelengths that appears 
to be very close to the lower limit given by the integrated light 
of resolved galaxies\cite{pozzetti}. 
The background flux at these wavelengths 
accordingly seems to be strongly dominated by the direct starlight 
from galaxies, thus excluding a large contribution from other 
sources -- in particular from the first stars formed\cite{santos}. 
This result also indicates that intergalactic space is more 
transparent to $\gamma$-rays than previously thought.
\end{abstract}

\phantom{}\vskip 0.05cm \noindent

\hspace{-1mm}The observations were carried out with the 
High Energy Stereoscopic System\cite{hess2155} (\hess),
a system of four imaging  atmospheric \C telescopes operating 
at energies $E\geq0.1$ TeV.
These two blazars are at present the most distant sources
for which spectra have been measured at these energies (Tab. 1).

Intergalactic absorption is caused by the process of photon-photon collision
and pair production. The original spectrum emitted by the source
(which we call ``intrinsic") is modified such that the observed flux 
$F_{\rm obs}(E)=F_{\rm int}(E)\cdot e^{-\tau(E)}$, 
where the optical depth $\tau(E)$ depends on the 
Spectral Energy Distribution (SED) of the 
Extragalactic Background Light (EBL) (Fig. 1).
Details are provided in the Supplementary Notes and Figures.
For any reasonable range of fluxes at ultraviolet (UV) and 
optical/near-infrared wavelengths (O--NIR),
$\tau(E)$ -- and thus absorption -- 
is larger at 1 TeV  with respect to 0.2 TeV.
This difference makes the observed spectrum steeper
(that is, $\Gamma_{\rm obs}>\Gamma_{\rm int}$, 
for a power-law model  ${\rm d}N/{\rm d}E\propto E^{-\Gamma}$)
The spectral change $\Delta\Gamma$=$\Gamma_{\rm obs}-\Gamma_{\rm int}$ 
scales linearly with the EBL normalization, and 
becomes more pronounced at larger redshifts. 
Thus more distant objects provide a more sensitive diagnostic tool.

In general,
if the intrinsic spectrum were sufficiently well known,
$\tau(E)$ -- and thus the EBL SED -- could be effectively measured
by comparing intrinsic with observed spectra.
Blazars, however, are characterized by a wide range of possible spectra,
and the present understanding of their radiation processes is not yet
complete enough to reliably predict their intrinsic $\gamma$-ray spectra.
But for these two sources,
with O--NIR fluxes at the level of the ``direct" estimates, 
the intrinsic spectra needed to reproduce the \hess data 
become extremely hard (that is, they have small values of $\Gamma$), 
at odds with the currently known blazar physics 
and phenomenology. 
This can be avoided by reducing the energy dependence 
of the optical depth, assuming either very low O--NIR fluxes 
(reducing $\tau$ at 1\,TeV)
or very high UV--O fluxes (increasing $\tau$ at 0.2 TeV).
The latter case, however, would require unreasonably 
high UV fluxes (as discussed later).
We can then derive an upper limit on the EBL
by requiring the intrinsic spectrum to be compatible 
with the present knowledge of blazars.

To determine such a limit, a plausible shape for the EBL SED is assumed.
As a reference shape in the 0.1-10 \m\ range we adopted the phenomenological 
curve used in refs\cite{aha01,aha03}, 
which is designed to be 
in general agreement with the EBL spectrum expected 
from galaxy emission\cite{prim01,dwek}.
This curve, labelled P1.0 in Figs 1 and 2, 
was originally normalized to match the ``direct" estimates
at 2.2 and 3.5\,\m\  (refs\cite{arendt,hauser,wright01}).
Here we leave its normalization as a free parameter,
scaling P1.0 by different factors P (labelled accordingly, 
the curve scaled by $0.45\times$ is ``P0.45")  down  
to the lower limit obtained 
by the resolved galaxy counts\cite{pozzetti} ($\sim$P0.4). 
To reproduce 
the excess around 1.5 \m\ claimed from the 
Infrared Telescope in Space (IRTS) data\cite{matsumoto} 
(also argued by ref\cite{cambresy}),
an additional {\it ad hoc} component was considered, 
labelled ``E$_{\rm NIR}$".
This feature is not expected from standard galaxy evolution models,
and could be the spectral signature of radiation produced 
in the early universe, for example by the first stars formed 
(metal-free massive stars, called  `Population III'; 
see e.g. refs\cite{santos,kashlinsky}).

The intrinsic source spectra (Fig. 2) 
have been reconstructed directly from the observed ones
using the assumed EBL,  without {\it a priori} assumptions on 
the blazar spectrum.
EBL evolution effects (e.g. due to galaxy evolution) were not included:
these effects, negligible at low redshifts, become important as redshift 
increases, but for the range considered here ($z=0.165-0.186$) their impact
is still limited to a factor of $\lesssim10$\% ($\Delta {\rm P}<0.1$; 
see Supplementary Information).
The reconstructed spectra are generally compatible with a power-law 
(${\rm d}N/{\rm d}E\propto E^{-\Gamma}$), 
but the EBL densities P1.0+E$_{\rm NIR}$ and P1.0 
both yield extremely hard photon indices ($\Gamma_{\rm int}<0$, 
see Fig. 2 and Supplementary Fig. SI-2),
implying a pile-up or line-like feature in the blazars' SED around 1--3\,TeV.
We obtain the same result by
considering the NIR excess added to the galaxy counts limits
(e.g. P0.4+E$_{\rm NIR}$, see Supplementary Fig. SI-3).
This is because a lower EBL flux only in the UV--O band
decreases the absolute values of $\tau$ but
increases the contrast between 0.2 and 1 TeV.

Such hard spectra have never been seen in the closest, less absorbed 
TeV blazars such as Mkn~421 and Mkn~501\cite{hegra,whipple,cat} 
($z\sim0.03$, $\Gamma_{\rm int}\approx1.5-2.8$ using the same EBL SEDs), 
and are difficult to explain  within  the  present
standard leptonic or hadronic scenarios\cite{aha01} for blazar emission. 
In shock acceleration models, 
the hardest index obtained for the accelerated particles
is $s=1.5$ (see e.g. ref\cite{drury}). 
In the case of protons interacting with ambient plasma, 
the resulting $\gamma$-ray spectrum has the same slope, $\Gamma_{\rm int}=1.5$. 
In the case of electrons, 
the spectrum of the $\gamma$-rays emitted through inverse Compton
scattering is expected to be steeper than $\Gamma_{\rm int}=1.5$ 
under most circumstances.
Only if radiative cooling is not effective
{\it and} the blazar Compton emission is wholly within the Thomson limit
-- unlikely at such high energies --
one finds $\Gamma_{\rm int}=(s+1)/2=1.25$.
We thus assume in the following discussion that the true average 
intrinsic spectrum was not harder than $\Gamma_{\rm int}=1.5$,
although later we also address the possibility of harder spectra.

To be compatible with $\Gamma_{\rm int}\geq1.5$,   
the EBL flux has to be scaled down to P0.45 
to explain both objects' data, with
1ES\,1101$-$232 providing
the most stringent constraints 
thanks to the better statistics at high energies and the larger redshift.
With a fixed EBL shape, there is a direct link between the normalization 
P and $\Gamma_{\rm int}$.   
The one-sigma statistical and systematic uncertainties 
on the \hess spectral measurement can then be translated to an 
equivalent uncertainty on P, $\Delta{\rm P}\simeq\pm0.15$ 
(see Supplementary Information).

This limit (P0.45) is robust with respect to a different EBL spectral shape, 
as long as it maintains an overall maximum around 1--2\,\m.
Below 1\,\m, lower fluxes than our template tend to 
harden the intrinsic spectrum more, 
while even a flat slope 
from 1.4\,\m\ down to 0.1\,\m\ would soften it by only $\Delta\Gamma\sim$0.1.  
Above 2\,\m, the slope cannot be much flatter than our template 
-- a flatter slope could explain in part the ``direct" measurements
at 3.5\,\m\  (Supplementary Fig. SI-4) --  
because this would again imply  
a new, very hard component ($\Gamma_{\rm int}<0$) in the intrinsic spectrum, 
rising at a few TeV (Supplementary Fig. SI-5).
In this respect, this \hess dataset  gives the same indication as
the HEGRA data\cite{aha03} on H\,1426+428 (z=0.129),
which show a flattening feature above 1 TeV  naturally provided 
by a starlight EBL between 3 and 10 \m\  (SED $\propto\lambda^{-1}$).

Therefore, the  
conservative and self-consistent assumptions of 
both  not unusual blazar spectra ($\Gamma_{\rm int}\geq1.5$) and 
a galaxy-like EBL spectrum
allow the EBL flux around 1--2\,\m\ to be constrained at the
level of $\lesssim(14\pm4)$\,\nw  (that is, $\leq 0.55\pm0.15$ $\times$P1.0).
This corresponds to P(0.45+0.1) to allow for galaxy evolution effects.
Coupled with the lower limits derived from
galaxy counts given by the Hubble Space Telescope\cite{pozzetti}
($\sim$$9.0$--$9.7^{+3.0}_{-1.9}$\,\nw), 
the \hess spectra lead to conclude that 
more than 2/3 of the EBL in the O--NIR band is resolved into single sources.
This result is completely independent of any ``direct" measurement
of the EBL. 
Remarkably, it is in severe conflict with the claims of high EBL flux 
at NIR wavelengths\cite{matsumoto,cambresy}
and, to a lesser extent, with the reported detections
at 2.2 and 3.5\,\m\cite{hauser,wright01}.  
The \hess upper limits agree instead with the most recent theoretical 
calculations\cite{prim05} of the EBL,
as well as with recent theoretical arguments\cite{madausilk,dwek2} 
against high EBL fluxes due to Pop III stars.

This result is also rather insensitive to
small changes in the assumed intrinsic slope.
A different value, if proved more likely by future results, 
will shift the limit accordingly, but only 
strong spectral differences would qualitatively change our conclusion:
even a value of $\Gamma_{\rm int}=1.0$ would loosen 
the $0.55\times$ limit only to $\leq0.7\times$P1.0.

Alternative scenarios which could reconcile the measured spectra 
with high O--NIR fluxes formally exist, and would
represent a major discovery in their own right,
but we consider them  very unlikely, given their exotic implications.
Higher UV fluxes would make the intrinsic spectra softer,
but the huge values required ($>300$ \nw, see Fig. 1, for example) are not 
supported by other measurements\cite{hauser,bernstein}, 
and could not be accomodated within any reasonable cosmological model.

A more viable  alternative is 
that such hard spectra are a real, new feature of the TeV blazar emission.
Possible mechanisms have already been envisaged\cite{aha01}. For example,
the inverse Compton scattering of mono-energetic electrons 
($E_0$, such as cold plasma with very large bulk motion Lorentz factor),
interacting in deep Klein-Nishina regime with 
a narrow-band photon field (such as a Planck-type distribution) 
may lead to very flat $\gamma$-ray spectra with a
sharp pile-up at $\epsilon_\gamma \sim E_0$, reproducing spectra like the 
ones in Fig. 2.  However,
such features should become directly visible in the 
observed spectra of the closer, less absorbed objects of the same type,
like the well-studied  Mkn 421 and Mkn 501
(if $\Gamma_{\rm int}$ was $\approx0$, 
they should show $\Gamma_{\rm obs}\lesssim0.5$).
This is in contrast with observations\cite{hegra,whipple,cat}, 
unless we assume a dependence of the source parameters on redshift 
such that the corresponding features always disappear due to EBL absorption.
It is difficult to justify such a fine-tuning on a relatively
small redshift range, although more objects and observations 
are needed to settle this issue definitively, 
given the still-limited sample.

Other possibilities include 
the non-cosmological origin of blazars' redshifts and 
the violation of Lorentz invariance (see e.g. ref\cite{proth}).
However, both scenarios imply dramatic revisions of modern physics and 
astrophysics, which we do not consider to be justified
by these data alone.

A low EBL level, in agreement with the expectations from standard galaxy 
evolution models,
is the simplest and most likely explanation of the \hess data.
\vskip 1.8cm

\begin{small}

Received 25 July 2005, accepted 22 February 2006.

\end{small}

\begin{addendum}
\item[Supplementary Information] is linked to the online version 
of the paper at www.nature.com/nature. 
\item The support of the Namibian authorities and of the University of Namibia
in facilitating the construction and operation of H.E.S.S. is gratefully
acknowledged, as is the support by the German Ministry for Education and
Research (BMBF), the Max Planck Society, the French Ministry for Research,
the CNRS-IN2P3 and the Astroparticle Interdisciplinary Programme of the
CNRS, the U.K. Particle Physics and Astronomy Research Council (PPARC),
the IPNP of the Charles University, the South African Department of
Science and Technology and National Research Foundation, and by the
University of Namibia. We appreciate the excellent work of the technical
support staff in Berlin, Durham, Hamburg, Heidelberg, Palaiseau, Paris,
Saclay, and in Namibia in the construction and operation of the
equipment.
\item[Competing interests statement] The authors declare that they have no
competing financial interests.
\item[Correspondence] \hspace*{-2mm}and requests for materials
should be addressed to L.C.~(email: Luigi.Costamante@mpi-hd.mpg.de).
\end{addendum}

\begin{table}
\begin{flushleft}
\end{flushleft}
\begin{center}
\begin{tabular}{lclclccl}
\vspace*{-1mm}\\
\hline
\vspace*{-3mm} \\
Source  & z  & Exp.  &  Sig.     & En. range & $\Gamma_{\rm obs}$ & N$_0$  &  \rchisq/d.o.f.  \\
       &     & (hrs) & ($\sigma$) & (TeV)   &          & (cm$^{-2}$\,s$^{-1}$\,TeV$^{-1}$)  &   \\
\hline  
\vspace*{-3mm} \\
1ES\,1101$-$232  &  0.186 & 43  & $\sim$12  & 0.16--3.3  & $2.88\pm0.17$ & $4.44\pm0.74\cdot 10^{-13}$  & 0.62/11 \\
H\,2356$-$309    &  0.165 & 40  & $\sim$10  & 0.16--1.0  & $3.06\pm0.21$ & $3.08\pm0.75\cdot 10^{-13}$  & 0.66/6 \\
\hline
\end{tabular}
\vspace*{9mm}
\caption{
{\bf Main parameters of the \hess observations}.
These observations were performed in June--December 2004 for H\,2356$-$309,
and March--June 2004 \& 2005 for 1ES\,1101$-$232. 
The table gives
the total exposure after selection for good quality data, significance 
of the detected $\gamma$-ray signal,
energy range used for the spectral fits and the result of a single power-law fit 
(${\rm d}N/{\rm d}E=N_0(E/\mathrm{TeV})^{-\Gamma_{\rm obs}}$).
The spectra have been calculated applying the technique described 
in ref\cite{hess2155}.
Errors are 1$\sigma$ statistical. The systematic uncertainty 
on the flux and photon index are estimated to be $\sim$15\% and $\sim$0.1, 
respectively.
Details of these observations will be published elsewhere;
here we have focused on the cosmological implications of the measured spectra.  
Compared to the previous observations of TeV blazars,
these \hess spectra provide significantly stronger constraints on the EBL 
because of the combination of a hard spectrum and relatively high redshift 
(see Supplementary Information).
}
\end{center}
\end{table}

\newpage

\begin{figure}  
\centering
\vspace*{-2.5cm}
\includegraphics[angle=0, width=15cm]{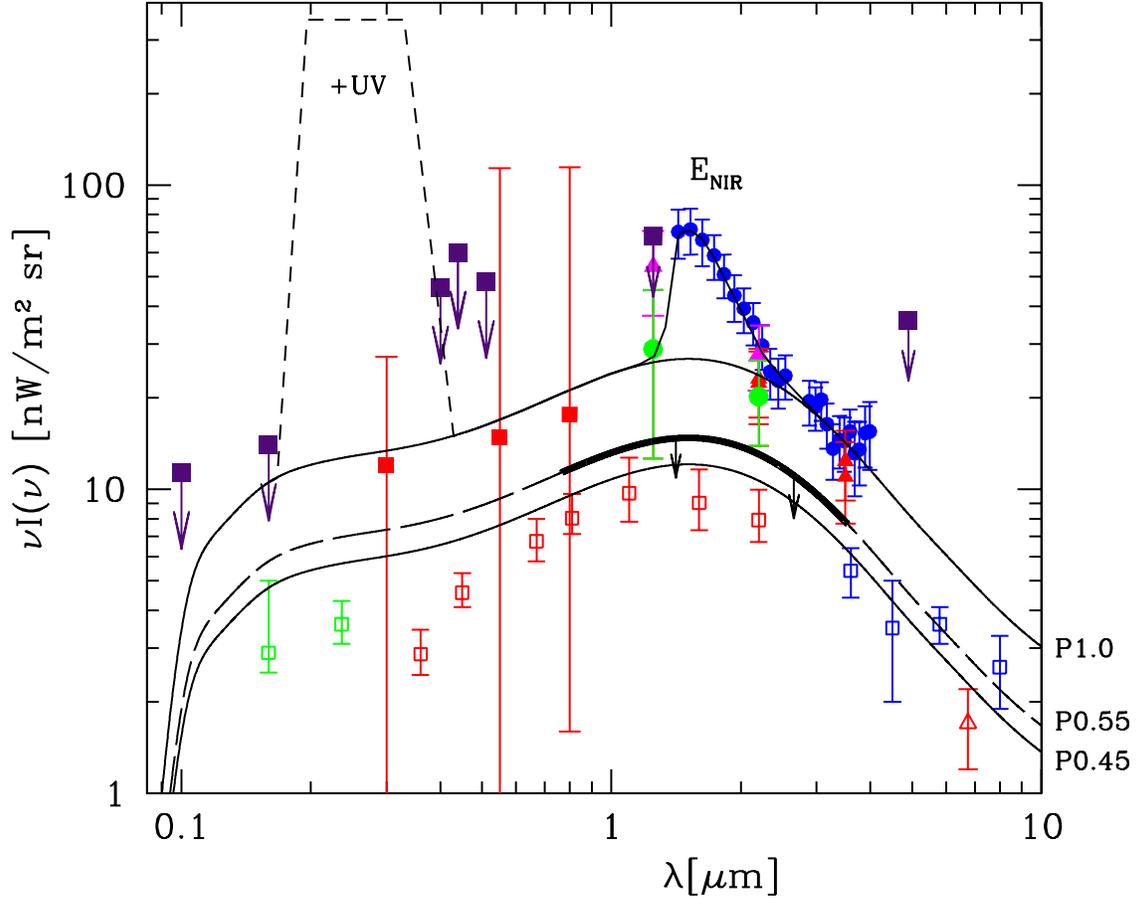}
\vspace*{-1.0cm}
\caption{ 
{\bf Spectral Energy Distribution of the EBL 
in the wavelength band most affecting these \hess data (0.1--10 $\rm \mathbf \mu m$)}.
The EBL data are from a review compilation\cite{hauser} 
(errors 1$\sigma$), unless otherwise stated.
Open symbols  correspond to the integrated light from galaxy counts,
and thus must be considered lower limits for the EBL:
in the UV--O range, from Hubble data (green, red\cite{pozzetti});
in the NIR, from Spitzer (blue\cite{fazio}) and ISO data. 
Note that these data are also lower limits for the total
emission from galaxies, because of various observational and selection effects
in the detection and counting of faint galaxies.
The possible missed light in the the UV--O band 
has been estimated\cite{totani} to be 
$\lesssim(2$--$3)$ \nw.
The upper limits (purple) are $2\sigma$ estimates\cite{hauser}. 
Direct measurements are shown as filled symbols:  
IRTS data from the NIR Spectrometer\cite{matsumoto} (blue), and  
data from  COBE/DIRBE (green\cite{wright01},  magenta\cite{cambresy} 
and red triangles).  
Red squares correspond to tentative detections in the optical\cite{bernstein} 
with corrections according to ref\cite{mattila}.
The curves show the EBL shapes used to reconstruct the intrinsic spectra. 
P1.0 gives 26, 23 and 14 \nw at 1.25, 2.2 and 3.5 \m, respectively.
The thick line shows the range most effectively constrained by the \hess data.
In the long-dashed regions, higher fluxes than P0.55
would not be in conflict, as long as the 
flux in the 1-3\,\m\ range is within or around the limit.
The short dashed line shows the additional UV component 
needed by P1.0 to soften the intrinsic spectrum down to $\Gamma=1.5$
(see Supplementary Fig. SI-5; E$_{\rm NIR}$ would require even higher fluxes). 
This example is the most energetically economic solution,
limited to the narrow range $\sim0.2-0.4$\,\m\  to have the maximum effect
on the $\gamma$-ray spectrum with the minimum UV flux and the minimum
impact on the overall attenuation.
}
\end{figure}

\newpage

\begin{figure}   
\centering
\includegraphics[angle=0, width=15cm]{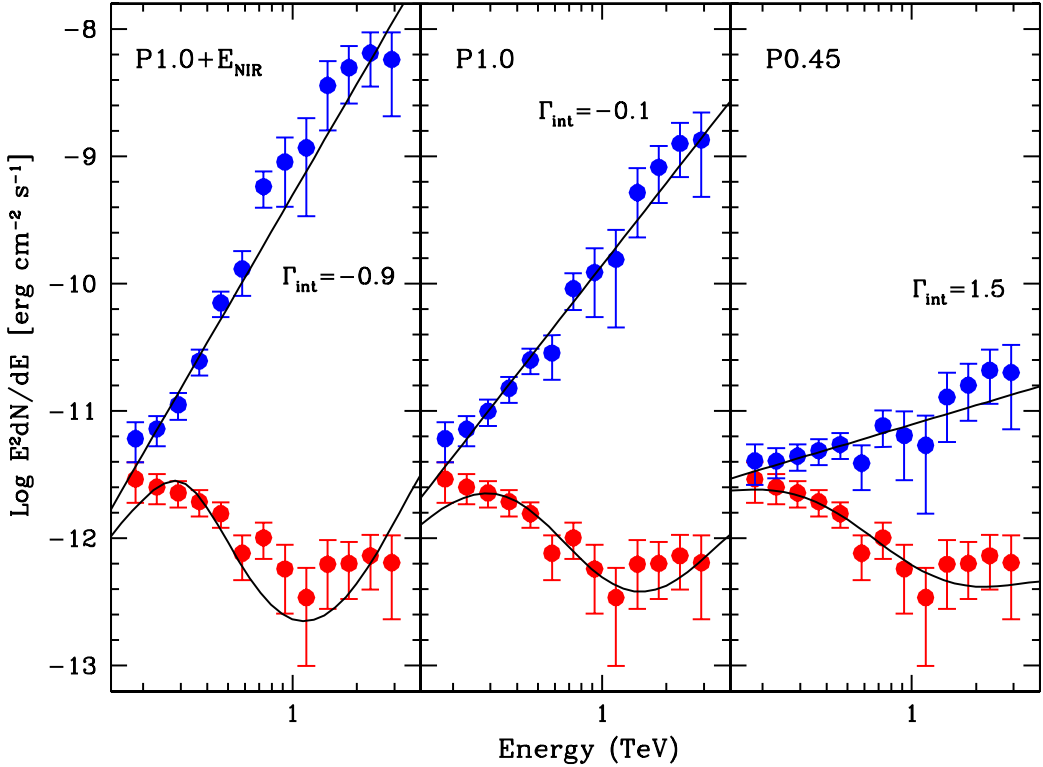}   
\vspace*{-6.0cm}
\caption{
{\bf The \hess spectra of 1ES\,1101$-$232, corrected for absorption 
with three different EBL SEDs, as labelled in Fig. 1}.
Red: observed data. Blue: absorption-corrected data.
The data points are at the average photon energy in each bin, 
also used to calculate the optical depth for reconstruction.
For the calculation, a  flat $\Lambda$-dominated cosmology was adopted,
with $H_0=70\;\rm km/s/Mpc$, $\Omega_m=0.3$, $\Omega_{\Lambda}=0.7$.
Error bars are 1$\sigma$, statistical errors only. 
Between 1.3 and 3.3 TeV, the overall detection is 4$\sigma$.
The lines show the best-fit power laws to the reconstructed spectrum
(${\rm d}N/{\rm d}E=N_0(E/\mathrm{TeV})^{-\Gamma_{\rm int}}$),
and the corresponding shapes after absorption.
The \rchisq/d.o.f.
(calculated by integrating the absorbed power-law model over each observed data bin)
are (from left to right): 1.20/11, 0.54/11, 0.47/11.
We note that possible spectral variability does not weaken our conclusions 
because it would imply states with even harder spectra than the average one (by definition).
Note also that the X-ray spectrum (which in blazars
usually samples the synchrotron emission of TeV electrons) measured 
during simultaneous observations in June 2004 and March 2005 does not show such hard slopes, 
but is similar to the historical values (Aharonian et al. 2006, in preparation).
For H\,2356$-$309, the same EBL SEDs yield 
$\Gamma_{\rm int}=-0.6$, $0.7$ and $2.0$, respectively (Suppl. Fig. SI-2).
The NIR excess onto the galaxy counts limits (P0.4+E$_{\rm NIR}$)
yields $\Gamma_{\rm int}\approx-0.7$ and $-0.4$ for 
the two objects, respectively (see Suppl. Fig. SI-3).
}
\end{figure}


\newpage


\noindent {\bf \LARGE Supplementary Information} \\

\noindent In this paper we have derived an upper limit on the Extragalactic Background Light
(EBL) in the optical to near infrared band (O--NIR).  
This is done by examining the effect of absorption, caused by interaction with EBL 
in this band, on the $\gamma$-ray spectra of two newly-detected distant blazars: 
1ES\,1101$-$232 (z=0.186) and H\,2356$-$309 (z=0.165).
Here we provide additional details
on the manner in which EBL absorption modifies the source spectrum, and 
on the impact on the derived limit
of the uncertainties in the $\gamma$-ray measurements and
of galaxy evolution effects.
We also explain why previous TeV blazar detections did not provide a comparably strong limit,
and present supplementary figures to better illustrate the cases
described in the paper. At the end, we provide a table with the 
\hess spectral data for the two sources.

\section{\Large Supplementary Notes}

\subsection{EBL absorption} \  \\ 
Absorption modifies the intrinsic source spectrum according to
$F_{\rm obs}(E)=F_{\rm int}(E)\cdot e^{-\tau(E)}$.
The optical depth $\tau(E)$ is obtained by convolving the 
EBL photon density distribution
$n(\epsilon)$ with the pair production cross-section $\sigma(E,\epsilon)$,
and integrating over the $\gamma$-ray photon path from the source 
to the observer:

\begin{equation}
\tau(E) = \int_{0}^{z} {\rm d}l(z) \int_{\epsilon_{\rm thr}}^{\infty} n(\epsilon) \sigma (E,\epsilon) {\rm d}\epsilon ,
\end{equation} 

\noindent (where  ${\rm d}l(z)=c\,{\rm d}t(z)$ is the proper line element,
$\sigma(E,\epsilon)$ is already integrated over 
angles for an isotropic field of background photons, 
and the photon density and energies depend on redshift). 
The energy dependence of $\tau(E)$ strongly depends on the spectrum
of the background photons (see e.g. ref[19]). 
For example, if $n(\varepsilon) \propto \varepsilon^{-1}$
(or equivalently the EBL SED $\nu I_\nu  \propto \lambda^{-1}$), the optical depth appears
constant with energy, while for  $n(\varepsilon) \propto \varepsilon^{-2}$
(i.e., $\nu I_\nu  =\rm constant$), the optical depth $\tau(E) \propto E$.
Note that in the first case, the observed spectrum simply reproduces 
the spectrum emitted by the source, attenuated by a constant factor. \\
The energy dependence of the optical depth translates to strong 
modifications of the source spectrum even with relatively small
differences in $\tau$, due to the exponential dependence.
For example, given an intrinsic power-law spectrum 
${\rm d}N/{\rm d}E\propto E^{-\Gamma_{\rm int}}$,
a difference of only one unit in $\tau$ between 0.2 and 1 TeV
increases the spectral index by $\Delta\Gamma=\Gamma_{\rm obs}-\Gamma_{\rm int}=0.62$.
Figure SI-1 shows the attenuation factor as a function of energy
in four  relevant cases (P1.0, P0.4, P0.4+E$_{\rm NIR}$, P1.0+UV). \\
For a $\gamma$-ray photon of a given energy $E$,
because of the threshold effect
($E\epsilon\geq (m_ec^2)^2$),
only EBL photons of energy above threshold (i.e., $\lambda<\lambda_{\rm thr}$)
contribute to the absorption.
Also, because of the relative narrowness of the 
$\gamma\gamma\rightarrow e^{+}e^{-}$ cross-section, 
for broad-band EBL spectra more than half of the interactions 
occur with a narrow interval of target photons 
$\Delta\lambda\sim(1\pm1/2)\lambda^\ast$ (ref[6,19])
around the cross-section peak, where 
$\lambda^\ast\approx1.4\;(E_{\gamma}/\,1 \, {\rm TeV})\, \mu{\rm m}$
(this relation is illustrated on the upper axis in Fig. SI-3 and SI-4,
as a guideline). 
Thus, the optical depth at $\sim0.2$ TeV is controlled only by the EBL 
fluxes at UV--O wavelengths (below 1~micron),
since NIR photons are below the energy threshold,
while the optical depth at $\sim1$ TeV is controlled mainly by the NIR flux.
The UV--O band has a more limited impact
on $\sim1$ TeV photons since the interactions occur far from the cross-section peak.
An increase or decrease of the UV--O flux alone
changes the optical depth at 0.2 TeV  much more than at 1 TeV. \\
Therefore, it is possible to reduce
the energy dependence of the optical depth (and thus to avoid the hard reconstructed spectra)
by  assuming either very low O--NIR fluxes (reducing $\tau$ at TeV energies) 
or very high UV--O fluxes (increasing $\tau$ at sub-TeV energies; 
e.g., see Fig. SI-1, right panel).

\subsection{Comparison with earlier observations} \  \\ 
These two new objects provide significantly stronger constraints on the EBL 
than any previous TeV blazar because of a favourable combination of 
hard observed spectrum and relatively high redshift. For sources
at smaller redshifts $\Delta\Gamma$ is smaller, while at the same redshift 
a softer spectrum can be misinterpreted as a hard spectrum 
which has been more heavily absorbed.
Both cases allow the EBL to have a wider range of possible fluxes
without yielding implausible values in the reconstructed blazar spectrum.
%
This is the reason why the TeV spectra of the two previous most 
distant objects,  1ES 1426+428 [11] 
(z=0.129, $\Gamma_{\rm obs}\simeq3.5\pm0.4$) 
and PKS 2155$-$304 [10] 
(z=0.116, $\Gamma_{\rm obs}\simeq3.3-3.4\pm0.06$), 
did not provide such strong constraints as these two new sources, even when measured 
with much higher signal-to-noise ratio (PKS\,2155$-$304).
The intrinsic spectra of H\,1426+428 and PKS 2155$-$304
start to become very hard ($\Gamma<1$) for EBL fluxes 
around P1.0+E$_{\rm NIR}$ and above.

\subsection{Statistical and systematics errors} \ \\
The determination of an EBL limit from gamma-ray spectra
needs to take into account the statistical and systematic errors
of the spectral determination, which amount for these data -- at the 1$\sigma$ level -- 
to $\Delta\Gamma\simeq0.2$ and 0.1, respectively.
In addition, the absolute energy scale of Cherenkov telescopes is only determined to $\sim 15$\%.
The EBL limit can be different, therefore, because for example the true blazar 
spectrum can be softer than the \hess value, and the photon energies lower.
Using the same $\Gamma_{\rm int}=1.5$ limit on the intrinsic spectrum, these errors 
can be translated into an equivalent uncertainty for the scaling factor P:
namely, 
by finding the values for which 
$\Gamma_{\rm int}\pm\Delta\Gamma_{\rm int,stat}\pm\Delta\Gamma_{\rm int,sys}=1.5$, 
and shifting the bin energies $\pm15$\%.
For the 1ES\,1101$-$232 spectrum, which provides the most stringent constraints, 
the errors in $\Gamma$ translate to $\Delta{\rm P}\simeq\pm0.1$
(the relation is linear, and is approximately $\Delta{\rm P}\simeq0.34\Delta\Gamma$),
and the energy scale error into $\Delta{\rm P}\simeq\pm0.05$.
All three contributions 
($\pm1\sigma$ statistical, $\pm0.1$ systematic, $\pm15$\% energy scale uncertainty)
combined yield $\Delta{\rm P}\simeq\pm0.15$.

\subsection{Galaxy evolution effects} \ \\
In the calculation of the optical depth,
we have not included the effects of galaxy evolution between $z$=$0$ and
$z=0.165$ to 0.186.  That is, 
we assume a constant (rather than decreasing with redshift) EBL co-moving energy density.
This corresponds to the assumption that all the background photons seen today
were already in place at the source redshift.
To estimate the impact of this approximation in the  
redshift range considered here, 
we used the model by Primack [12] in 2001,
interpolating a set of EBL SEDs as a function of redshift 
(kindly provided by the authors).
Including galaxy evolution, the intrinsic spectra become 
$\Delta\Gamma\lesssim0.2$ softer than without evolution, 
thus loosening the limit on the scaling factor P by less than 0.1. 
Note, however, that this effect can still be safely ignored 
when testing scenarios where the EBL is dominated by additional radiation
produced in the early universe (for which our assumption fully holds). 

\newpage
\setcounter{figure}{0}
\setcounter{table}{0}
\renewcommand{\thetable}{SI-\arabic{table}}

\begin{sfigure}   
\centering
\resizebox{\hsize}{!}{
\includegraphics[viewport=30 50 530 700,clip,width=12cm]{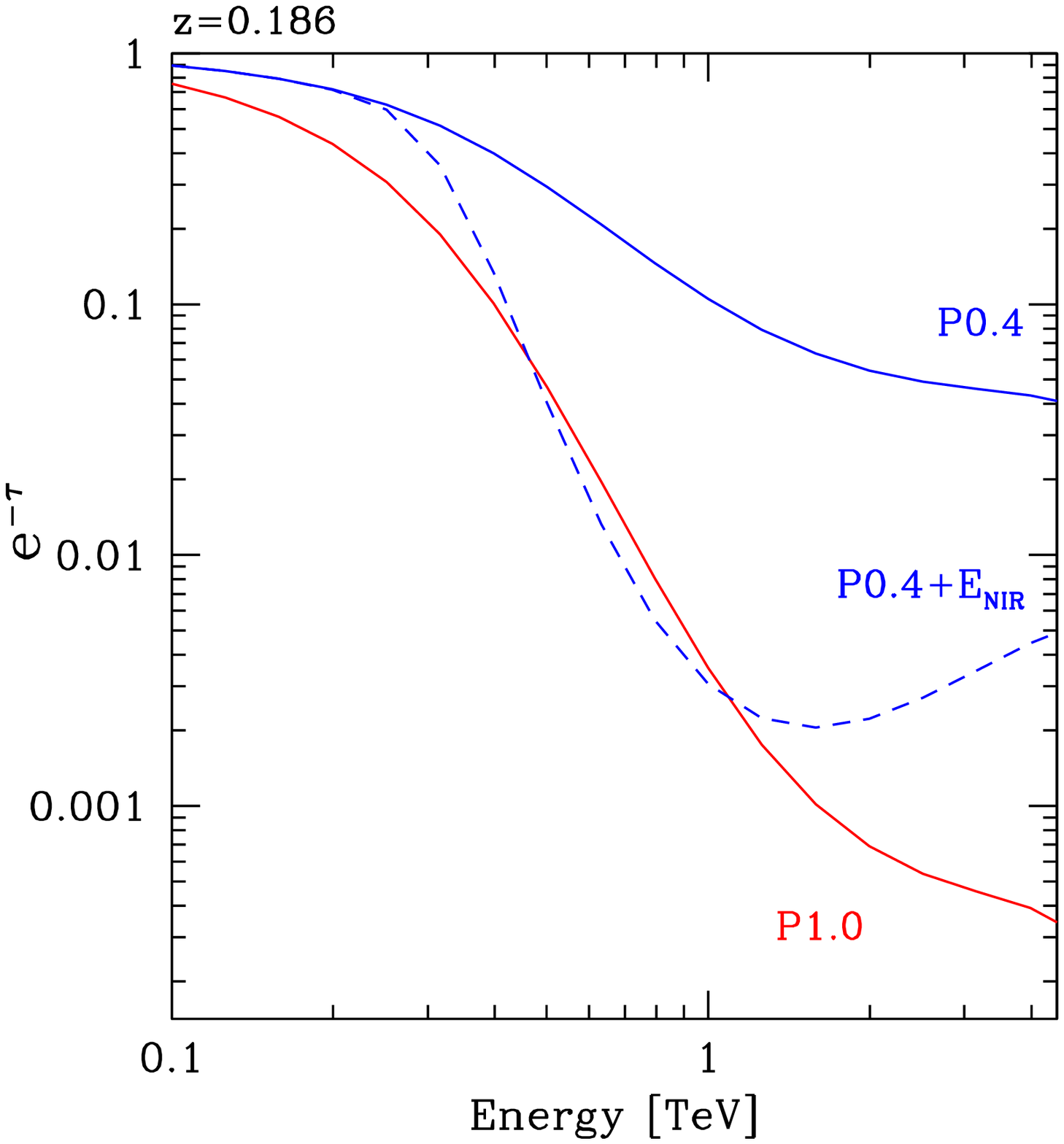}  
\includegraphics[viewport=30 50 530 700,clip,width=12cm]{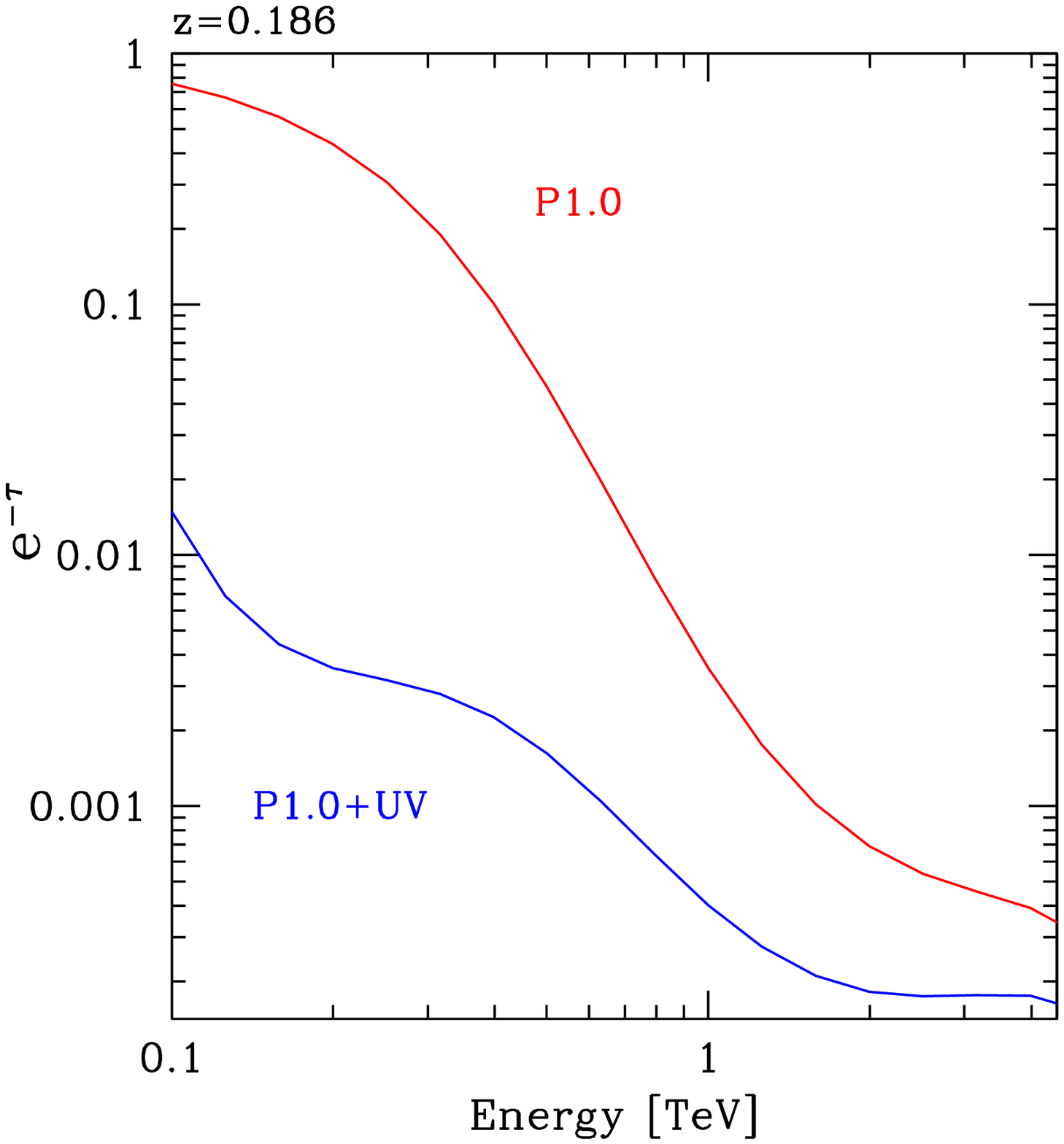} }
\caption{\small 
Attenuation factor $e^{-\tau}$ for the redshift of 1ES\,1101-232 (z=0.186),
assuming different EBL SEDs. Left panel: 
P1.0, P0.4, P0.4 with E$_{\rm NIR}$ added on (Fig. SI-3).
Right panel: P1.0 and P1.0+UV.
These curves would correspond to the observed TeV spectrum
for a source with a flat power-law emission spectrum
(e.g., for a source spectral energy distribution $E^2{\rm d}N/{\rm d}E$ with $\Gamma_{\rm int}=2$).
Above 0.2 TeV, the increase of $\tau$ with $E$
steepens the spectrum, up to $1$--$2\;{\rm TeV}$. 
Left panel: the thick lines (corresponding to P0.4, P1.0) show how 
an increase of the EBL normalization 
steepens the $0.2$--$1\;{\rm TeV}$ spectrum.
The dashed line (P0.4+E$_{\rm NIR}$) shows how the addition of a NIR excess 
around $\sim1$\m\ dramatically steepens the spectrum, due to 
the increase of $\tau$ for $0.8$--$1\;{\rm TeV}$ photons,
while photons at $0.1$--$0.3\;{\rm TeV}$ remain unaffected.
Above 2 TeV, the absorption curve corresponding to P0.4+E$_{\rm NIR}$ even hardens
with energy since P0.4+E$_{\rm NIR}$ is characterized by a slope steeper than $\lambda^{-1}$
between 2 and 6\,\m.
The right panel shows how an increase in the UV flux reduces the energy dependence
of the optical depth, at the price of a higher attenuation overall.
}
\end{sfigure}

\newpage

\begin{sfigure}   
\centering
\includegraphics[angle=0, width=14cm]{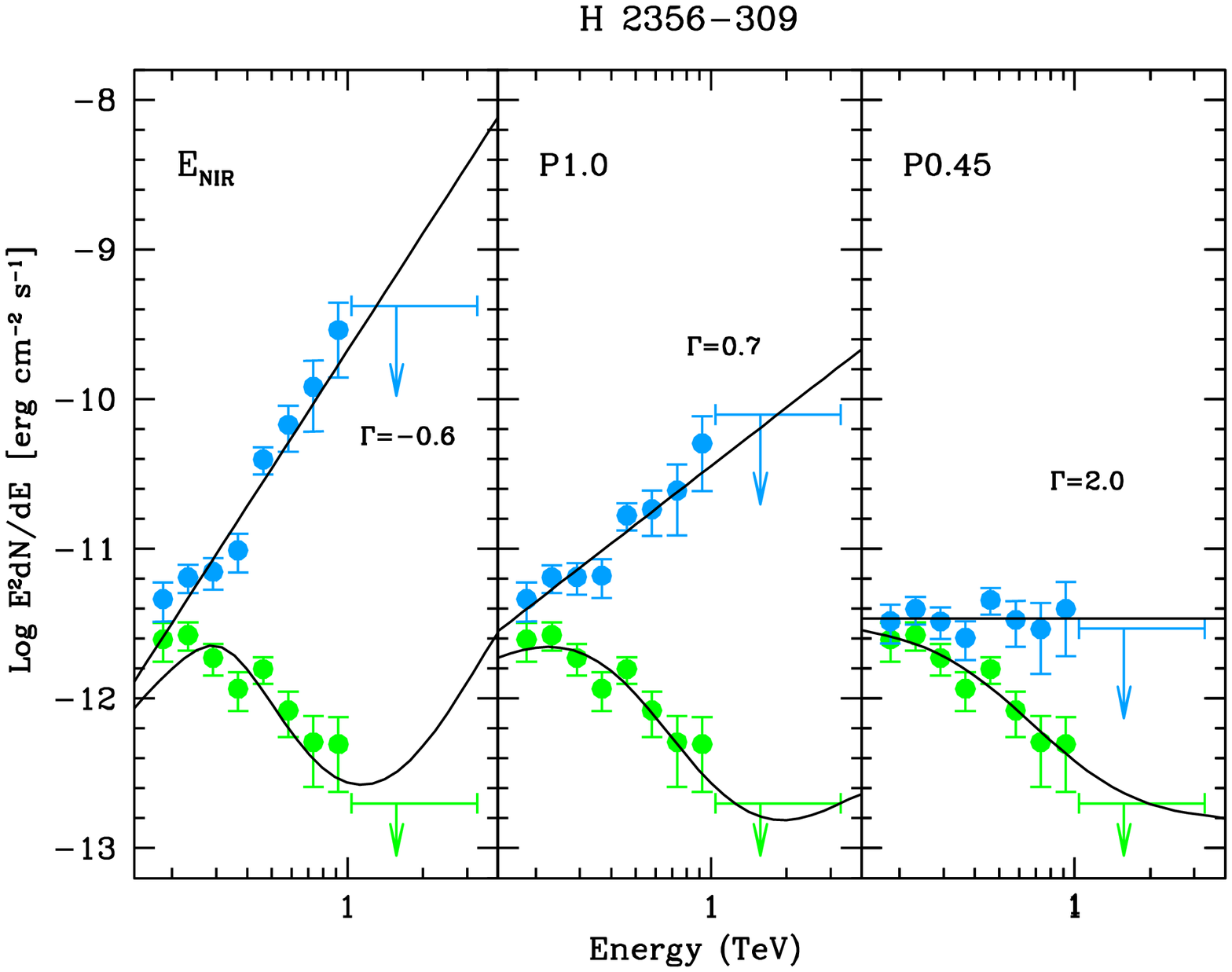}   
\caption{\small 
The \hess time-averaged spectrum of  H\,2356$-$309 (green),
and the absorption-corrected data (blue) with the same EBL SEDs as for 
1ES\,1101$-$232 (Fig. 1 and 2).
The data points and the 99\% upper limit
are at the average photon energy for each bin, also used to calculate
the optical depth for reconstruction.
Error bars are 1$\sigma$ statistical errors. 
The lines show the best-fit power laws to the reconstructed spectrum
(${\rm d}N/{\rm d}E=N_0(E/\mathrm{TeV})^{-\Gamma_{\rm int}}$),
and the corresponding shapes including absorption.
The \rchisq/d.o.f. are (from left to right): 
1.94/6, 0.86/6, 0.59/6.
}
\end{sfigure}

\newpage

\begin{sfigure}   
\centering
\resizebox{\hsize}{!}{\includegraphics[width=14cm]{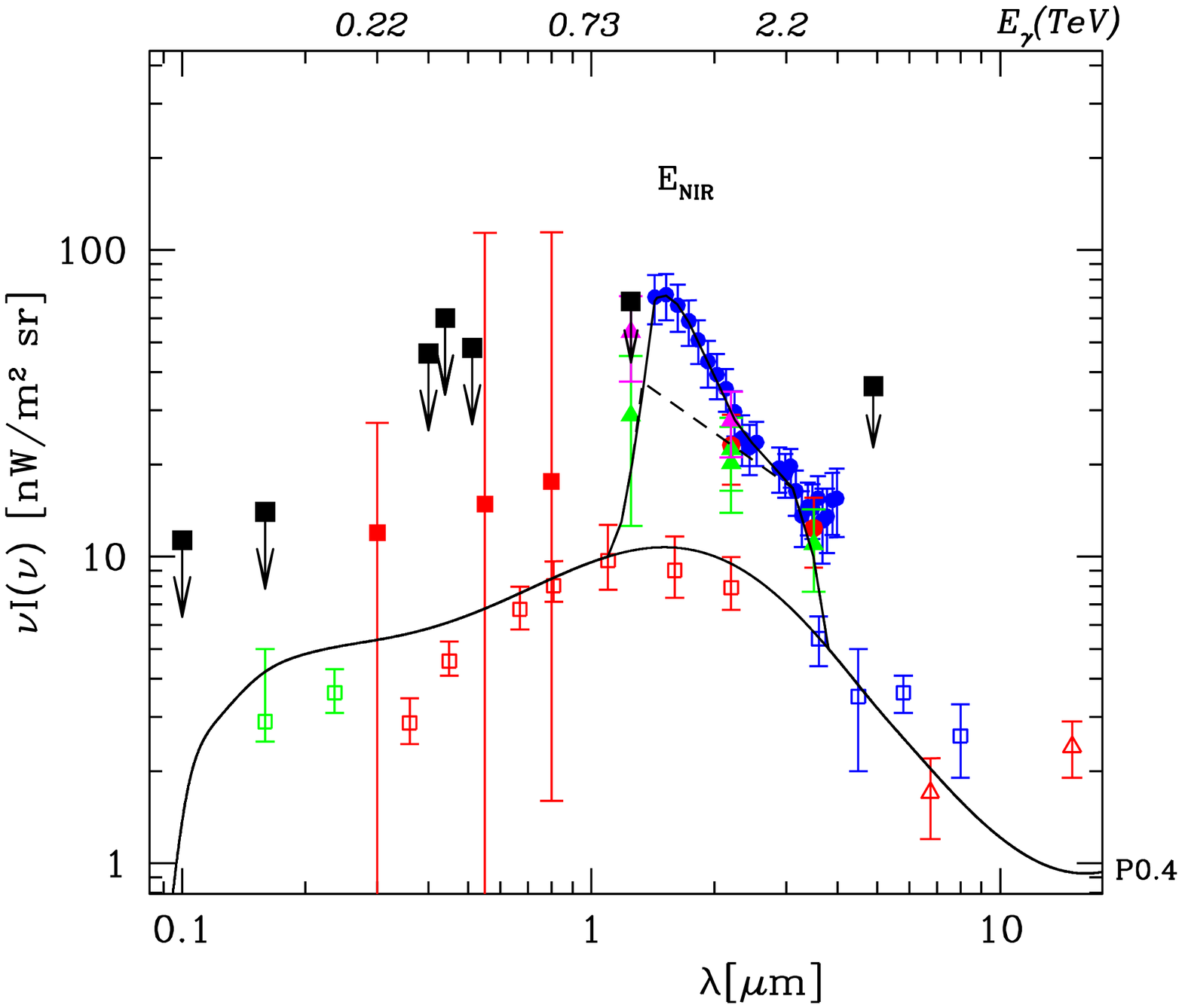}
\includegraphics[viewport=50 5 500 700,clip,width=10.1cm]{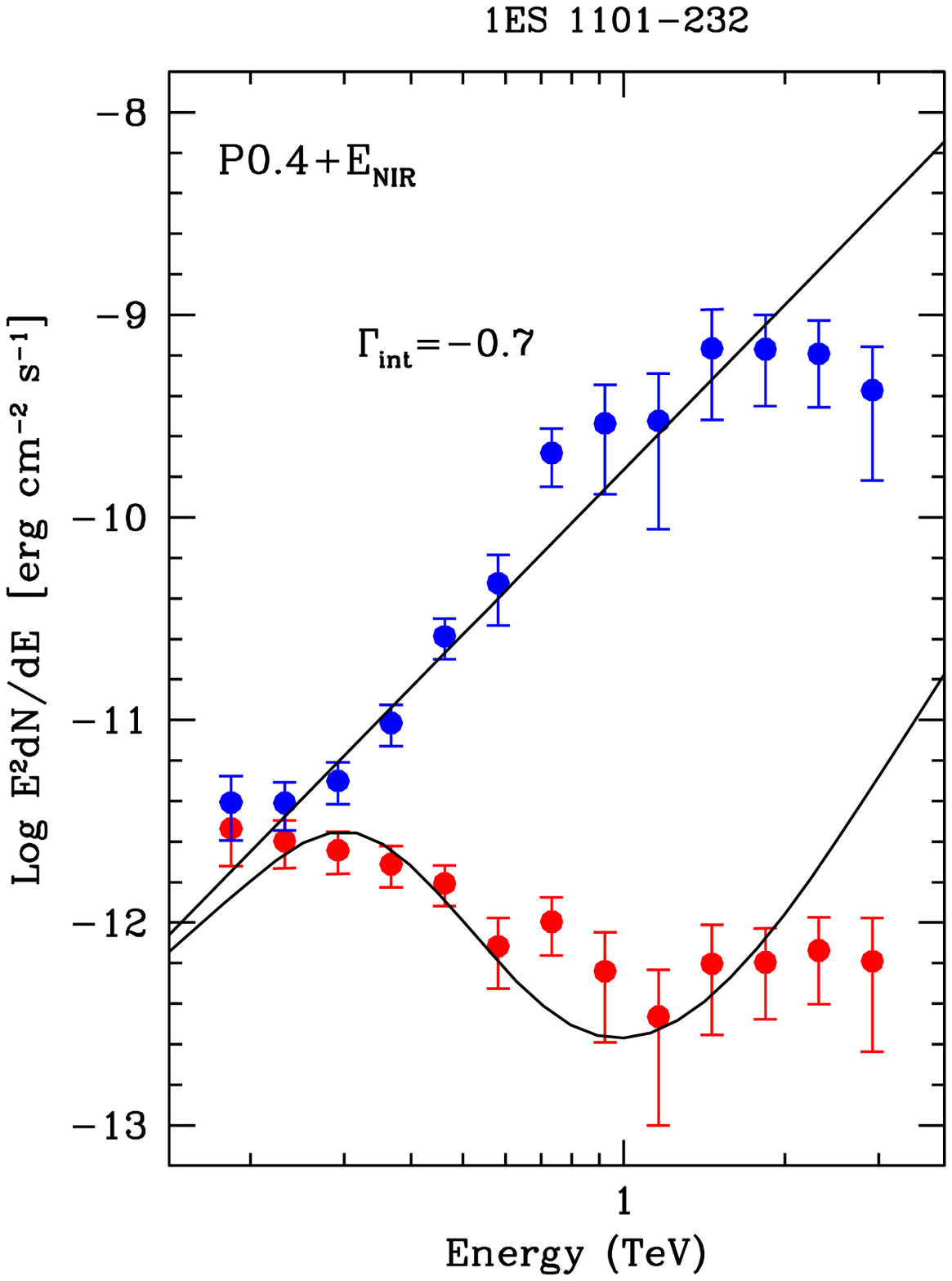} }
%
\caption{\small  
Effect of the NIR excess onto the galaxy counts limits.
Left:  EBL SED with the same data as Fig. 1. 
The full line  
shows the EBL shape with the NIR excess added on to P0.4,
while the right panel shows the corresponding absorption-corrected $\gamma$-ray spectrum
for 1ES\,1101$-$232.
Also, for this EBL SED, the intrinsic spectrum is extremely hard
($\Gamma_{\rm int}=-0.7$ between 0.1 and 2 TeV),
with a very sharp rise between 0.3 and 1.5 TeV where the intrinsic slope becomes as hard as 
$\Gamma_{\rm int}=-1.2$. This is due to the strong difference in optical depths 
at these two energies,
because the gamma-ray photons below 0.3 TeV are insensitive to the
NIR excess $>1$\,\m\  (since they are below the energy threshold for the
$\gamma\gamma\rightarrow e^{+}e^{-}$ process), 
whereas $\sim 1\,{\rm TeV}$ photons are fully affected since the interaction takes place
near the peak of the $\gamma\gamma\rightarrow e^{+}e^{-}$ cross-section.
The exclusion of the the IRTS data points between 1.4 and 2.2 \m\  (dashed line)
is not enough to avoid very hard spectra ($\Gamma_{\rm int}\approx0.2$):
significantly lower NIR fluxes are needed, as described in the main text.
We note that  ref[25] 
argues that the shape of the NIR excess between 1 and 4\,\m\ is almost identical to
the spectrum of the zodiacal light, and could therefore be of 
local instead of extragalactic origin.
}
\end{sfigure}

\newpage

\begin{sfigure}   
\centering
\vspace*{-3cm}
\includegraphics[angle=0, width=10cm]{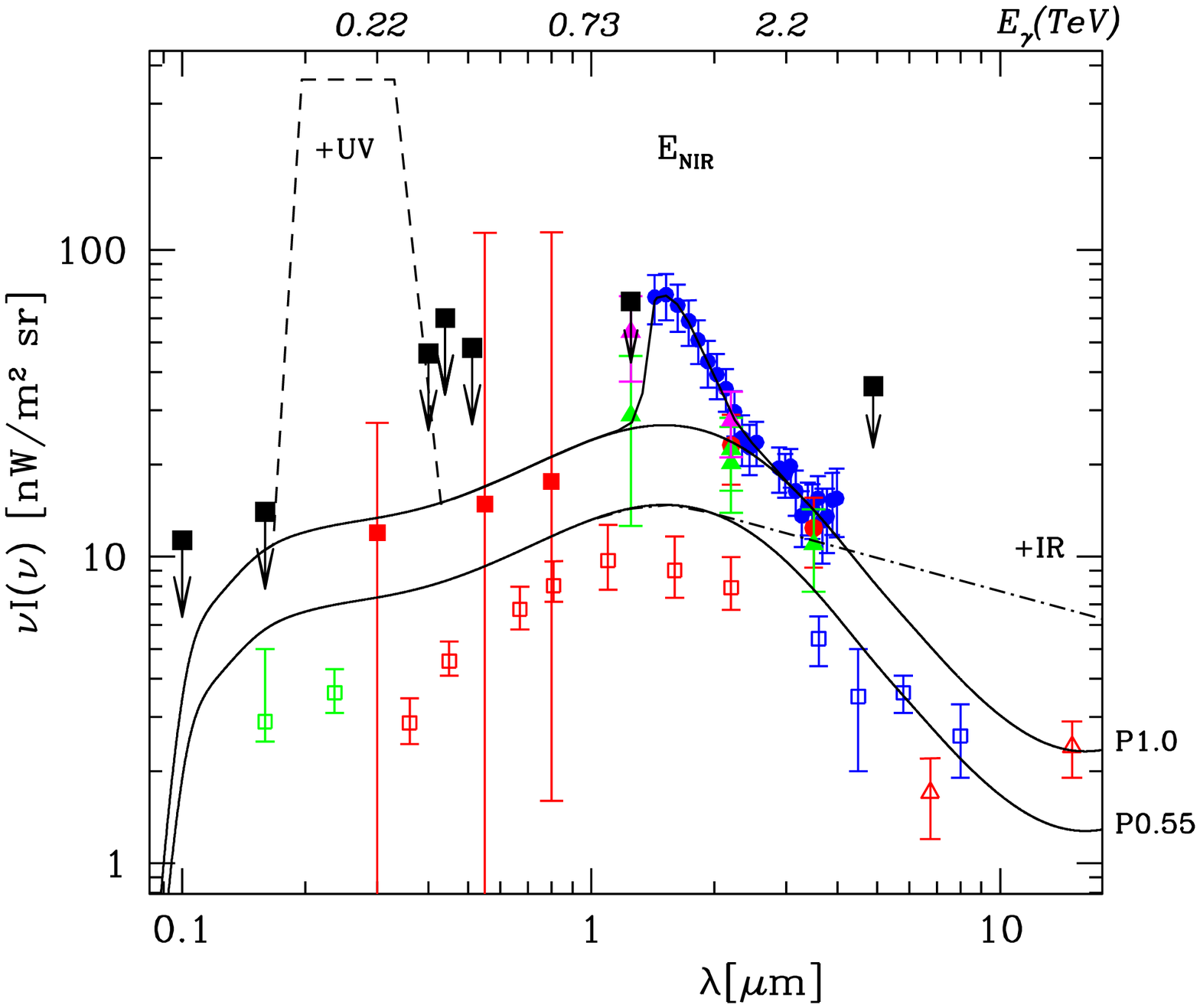}
\vspace*{-1cm}
\caption{\small  
Effect of modifications to the EBL SED in the UV and IR bands,
for the same data as Fig. SI-2. 
The short dashed line shows the additional UV component 
needed by P1.0 to soften the intrinsic spectrum down to $\Gamma_{\rm int}=1.5$,
as shown in Fig. SI-2. 
This is the most energetically-economic solution,
limited to the narrow range $\sim0.2$--$0.4$\,\m\  to have the maximum effect
on the $\gamma$-ray spectrum with the minimum UV flux and  overall attenuation.
The inclusion of E$_{\rm NIR}$ would require even higher UV fluxes, 
in the case of both  P1.0+E$_{\rm NIR}$ and P0.4+E$_{\rm NIR}$.
Above 2\,\m, a slope much flatter than our template (e.g. +IR, dot-dashed line)
is not supported by 
either the \hess (see Fig. SI-5) or HEGRA results [11], 
which instead suggest an EBL spectrum around $\propto\lambda^{-1}$.
If this is the case, it would constrain the EBL to be near the recent
Spitzer source counts [28], 
meaning that we are close to resolving the diffuse background also 
up to $\sim8$\,\m. 
}
\end{sfigure}


\begin{sfigure}   
\centering
\vspace*{-0.9cm}
\includegraphics[angle=0, width=9cm]{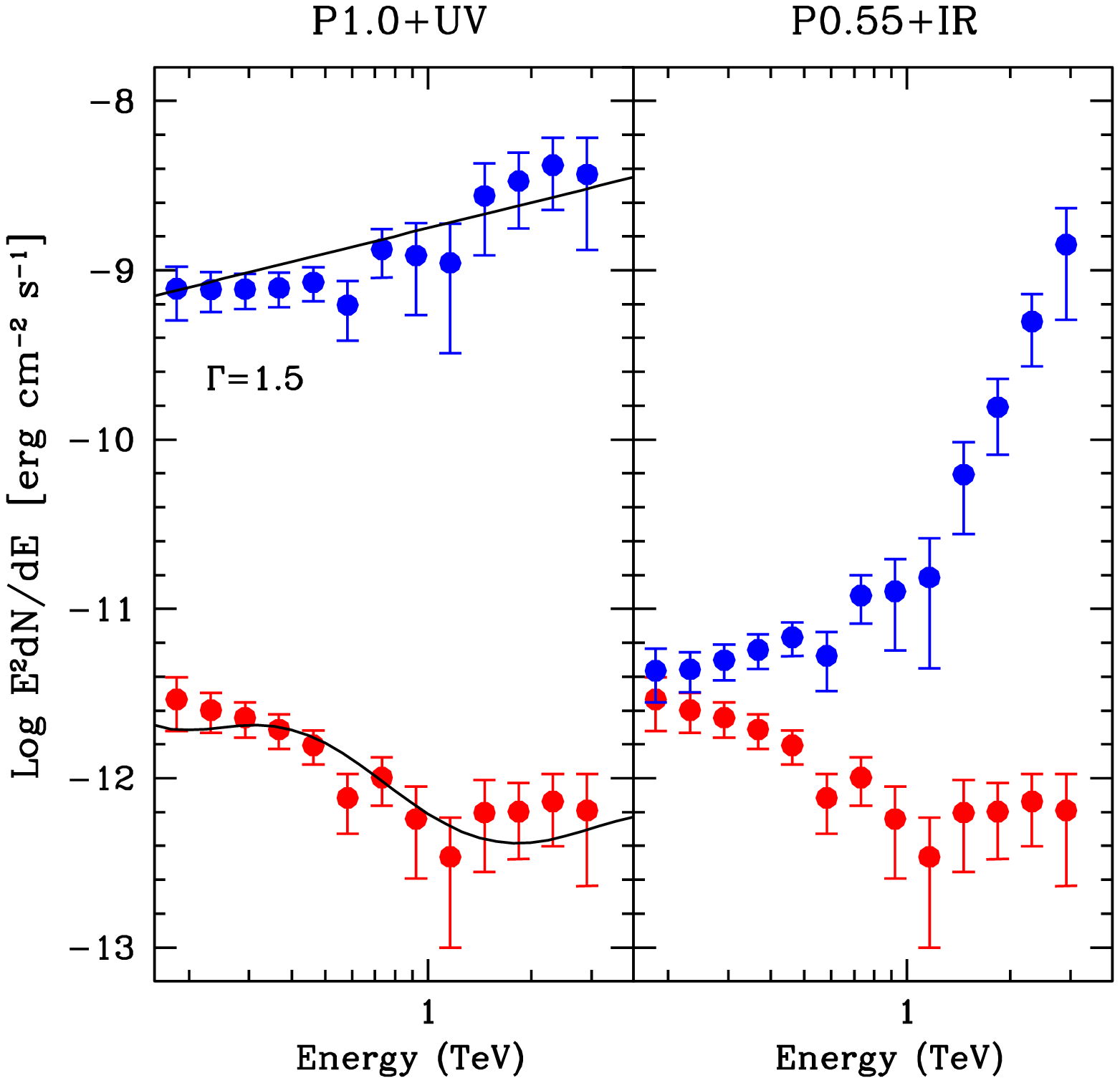}   
\vspace*{-1cm}
\caption{\small 
The \hess spectrum of 1ES\,1101$-$232, 
corrected for absorption with the EBL SED shown in Fig. SI-4.
The P1.0$+$UV panel shows 
the spectral softening given by the additional UV component, 
at the price of increased energetic requirements:
the intrinsic TeV flux would be 3 orders of magnitude higher than that observed,
and $\sim1.5$--$2$ orders of magnitude
above the blazar synchrotron peak, as indicated by simultaneous X-ray data
(Aharonian et al. 2006, in preparation).
The P0.55+IR panel 
shows the hardening given by 
an EBL spectrum above 2\m\
which is flatter than our template (Fig. SI-4, dot-dashed line).
%
%
}
\end{sfigure}

\begin{table}
\centering
\setlength{\tabcolsep}{4mm}
\caption{\hess spectral data. Tha table gives the
differential flux $\Phi$ in different energy bins, for both objects.
E$_{\rm avg}$ is the average photon energy energy in each bin. 
The upper limit for H\,2356-309 is given 
for the overall energy range [$1.041-3.292$] TeV,
with a confidence level of 99\%.}
\vspace*{1cm}
\begin{tabular}{cc|cc|cc}\hline
\multicolumn{1}{c}{Energy Bins}& & \multicolumn{2}{c|}{1ES\,1101-232}   & \multicolumn{2}{c}{H\,2356-309} \\
 E$_{\rm low}$ $-$ E$_{\rm high}$&  & E$_{\rm avg}$ & $\Phi\pm\Delta\Phi$    & E$_{\rm avg}$ &   $\Phi\pm\Delta\Phi$  \\
    $[$TeV$]$   &     & $[$TeV$]$&\multicolumn{1}{c|}{$[$\fu$]$}            & $[$TeV$]$  &\multicolumn{1}{c}{$[$\fu$]$}\\
\hline 
0.165 $-$ 0.208 &   & 0.184   &  5.38 $\,\pm$1.88  $\,\times\,10^{-11}$ & 0.184 &  4.58 $\,\pm$1.34  $\,\times\,10^{-11}$	 \\
0.208 $-$ 0.262 &   & 0.232   &  2.93 $\,\pm$0.78  $\,\times\,10^{-11}$ & 0.232 &  3.08 $\,\pm$0.66  $\,\times\,10^{-11}$   \\
0.262 $-$ 0.329 &   & 0.292   &  1.67 $\,\pm$0.39  $\,\times\,10^{-11}$ & 0.292 &  1.37 $\,\pm$0.33  $\,\times\,10^{-11}$   \\
0.329 $-$ 0.414 &   & 0.367   &  9.00 $\,\pm$2.08  $\,\times\,10^{-12}$ & 0.367 &  5.40 $\,\pm$1.57  $\,\times\,10^{-12}$   \\
0.414 $-$ 0.522 &   & 0.462   &  4.57 $\,\pm$1.04  $\,\times\,10^{-12}$ & 0.462 &  4.60 $\,\pm$0.94  $\,\times\,10^{-12}$   \\
0.522 $-$ 0.657 &   & 0.582   &  1.41 $\,\pm$0.54  $\,\times\,10^{-12}$ & 0.582 &  1.54 $\,\pm$0.52  $\,\times\,10^{-12}$   \\
0.657 $-$ 0.827 &   & 0.733   &  1.17 $\,\pm$0.37  $\,\times\,10^{-12}$ & 0.732 &  5.94 $\,\pm$2.96  $\,\times\,10^{-13}$   \\
0.827 $-$ 1.041 &   & 0.922   &  4.22 $\,\pm$2.34  $\,\times\,10^{-13}$ & 0.921 &  3.64 $\,\pm$1.89  $\,\times\,10^{-13}$   \\
1.041 $-$ 1.311 &   & 1.161   &  1.59 $\,\pm$1.13  $\,\times\,10^{-13}$ &  	    &	 \\
1.311 $-$ 1.650 &   & 1.462   &  1.83 $\,\pm$1.02  $\,\times\,10^{-13}$ &    1.572  &   $<$5.03 $\,\times\,10^{-14}$  \\    	  
1.650 $-$ 2.077 &   & 1.840   &  1.17 $\,\pm$0.56  $\,\times\,10^{-13}$ &       &	 \\
2.077 $-$ 2.615 &   & 2.316   &  8.49 $\,\pm$3.87  $\,\times\,10^{-14}$ &       	    &	 \\
2.615 $-$ 3.292 &   & 2.916   &  4.73 $\,\pm$3.03  $\,\times\,10^{-14}$ & 	   	    &	  \\
\hline
\end{tabular}
\end{table}

\end{document}